\documentclass[a4paper,superscriptaddress,aps,prl,twocolumn,floatfix,citeautoscript]{revtex4-1}
\usepackage[utf8]{inputenc}
\usepackage{amsmath}
\usepackage{graphicx}
\usepackage{cleveref}
\usepackage{physics}
\usepackage{xcolor}
\usepackage{cancel}
\usepackage{booktabs}
\usepackage{comment}
\usepackage[normalem]{ulem}

\newcommand{\UWMadison}{Department of Materials Science and Engineering, University of Wisconsin-Madison, 53706, USA}
\newcommand{\physics}{Department of Physics, University of Wisconsin-Madison, 53706, USA}
\newcommand{\chemistry}{Department of Chemistry, University of Wisconsin-Madison, 53706, USA}

\ifdefined\pdfpagewidth \pdfpagewidth=\paperwidth \pdfpageheight=\paperheight \fi
\AtBeginDocument{%
  \setlength{\oddsidemargin}{\dimexpr (\paperwidth-\textwidth)/2 - 1in \relax}%
  \setlength{\evensidemargin}{\oddsidemargin}%
}

\begin{document}
\title{Open-quantum-system theory of non-Markovian electron–phonon dynamics}

\author{Gabriele Riva}
\email{griva@wisc.edu}
\affiliation{\UWMadison}
\author{Jacopo Simoni}
\affiliation{\UWMadison}
\author{Yuan Ping}
\email{yping3@wisc.edu}
\affiliation{\UWMadison}
\affiliation{\physics}
\affiliation{\chemistry}

\begin{abstract}
We present a non-Markovian open-quantum-system formalism for nonequilibrium electron–phonon dynamics: a closed set of five coupled equations of motion, derived directly from the density matrix, in which memory, dissipative broadening, and coherent-phonon dynamics arise on an equal footing without two-time correlators. It recovers the Fan–Migdal, RPA-polarization, and Ehrenfest self-energies and the Lindblad and Boltzmann limits. In the strongly driven Holstein dimer it separates the coherent polaronic peak, of Ehrenfest origin, from the linewidths and multiphonon structure of Fan–Migdal origin.
\end{abstract}
%

\maketitle

Electron–phonon interactions govern a wealth of fundamental phenomena in condensed matter and materials physics, from charge-density-wave instabilities~\cite{Rice75,cao2023} and polaron formation~\cite{Berciu2006,Giu2022,Esterlis2018} to ultrafast spin and carrier relaxation and energy dissipation following photoexcitation~\cite{Kli00,oth98,Xu21,Xu2024-cb}.
Despite decades of progress, a fully quantum, non-perturbative treatment of nonequilibrium electron–phonon coupling—one capturing quantum coherence, memory effects, and dissipative dynamics of both electronic and phononic degrees of freedom—remains a formidable challenge~\cite{Stef_book, Giu2017,Andre_book,Perf2023,Melo2016,Pav2022,vlcek2024,Karl18,Rob22I,Rob22II,Reev2023,stef24,Mar13}.

A natural and powerful framework for addressing this challenge is provided by the theory of open quantum systems~\cite{Petru_book,Xu21,Rossi2002, Iotti_2017}, in which the electronic subsystem is regarded as an open system coupled to a phonon environment. Within this picture, the central object of interest is the one-body reduced density matrix (1-RDM), whose equation of motion (EOM) encodes the full single-particle dynamics of the electrons. The phonon environment enters through its coupling to the electrons, described by an EOM for the electron-phonon correlations. Crucially, it is the coupled system of EOMs for the electronic 1-RDM and the electron-phonon correlations that generate memory effects in the electronic dynamics: the time evolution of the 1-RDM depends explicitly on its own history, with the electron-phonon correlation function acting as a memory kernel that retains information about the past state of the coupled system. Such non-Markovian effects are of paramount importance in the ultrafast regime~\cite{bonitz2016} — relevant, for instance, to pump-probe spectroscopy and photoinduced phase transitions \cite{PhysRevB.97.165416,PhysRevLett.123.097601} — where the phonon bath cannot be assumed to relax instantaneously and where the conventional Markov approximation breaks down~\cite{Perspective}. At longer timescales, the same framework naturally describes the irreversible dissipation and relaxation of electronic energy into the lattice.

A particularly rich manifestation of electron-phonon physics arises in the polaron problem, in which an electron becomes dressed by a coherent cloud of lattice distortions~\cite{Robert2015,PavII2022}. The phononic degrees of freedom enter the formalism through three physically distinct equations of motion. The \textit{first} is the evolution of the phonon reduced density matrix, whose diagonal (off-diagonal) elements describe the phonon number beyond the thermal-average limit (phonon decoherence and lifetime). The \textit{second} captures the anomalous, two-phonon correlator, associated with pair creation and annihilation processes. The \textit{third} is dedicated to the coherent phonon~\cite{Stef25} — the classical or macroscopic part of the phonon field, characterized by a nonzero expectation value of the phonon displacement — which plays a decisive role in polaron formation and dynamics, as well as in the response to ultrashort optical pulses that resonantly drive coherent lattice motion~\cite{Wri1992,Loud2001,Mats2004}.


In this Letter, we present a unified, non-Markovian open-quantum-system formalism for electron–phonon coupled systems that treats coherent and dissipative dynamics on an equal footing, 
capturing quantum coherence, memory effects, dissipative broadening, and coherent-phonon dynamics within a single framework.
%
This framework captures dissipation and relaxation beyond the Markov and weak-coupling limits while simultaneously accounting for coherent phonon dynamics and polaron formation.
We first introduce the 1-RDM theory and establish its link to nonequilibrium Green's function (NEGF) theory, and then validate it against the exact solution of the Holstein dimer.

We consider the following Hamiltonian
\begin{equation}
    \hat{H}(t) = \hat{H}_{\rm e} +  \hat{H}_{\rm ph} + \hat{H}_{\text {e-ph}} + \hat{H}_{\rm env}(t) \,,
\end{equation}
with
\begin{align}
    &\hat{H}_{\rm e} = \sum_1 \varepsilon_1 \hat{c}_1^\dagger \hat{c}_1 \nonumber\\
    &\hat{H}_{\rm ph} = \sum_{\bf q} \omega_{\bf q} \hat{b}_{\bf q}^\dagger \hat{b}_{\bf q} \nonumber\\    &\hat{H}_{\rm env} (t) = \sum_{12} v^{\rm ext}_{12}(t) \hat c^\dagger_1 \hat c_2 \nonumber\\
    &\hat{H}_{\text {e-ph}} = \sum_{\bf q}\sum_{1,2}\big(g_{12}^{{\bf q}-}\hat{c}_1^\dagger \hat{b}_{\bf q}\hat{c}_{2} + g_{12}^{{\bf q}+} \hat{c}_{2}^\dagger \hat{b}_{\bf q}^\dagger \hat{c}_1\big)\ \,,\label{eq:sys-env}
\end{align}
where $\varepsilon_1$ and $\omega_{\bf q}$ are the non-interacting energies of the electrons and phonons, respectively. $\hat c$ ($\hat c^{\dagger}$) denotes annihilation (creation) operator for electrons, $\hat b$ and $\hat b^{\dagger}$ the same but for phonons. $v_{12}^{\rm ext}(t)$ is the time dependent interaction with an external electromagnetic field used to drive the electronic system out of equilibrium~\cite{Perspective}, and $g_{12}^{\bf q \pm}$ are the absorption ($+$) and emission ($-$) electron-phonon coupling matrix~\cite{Giu2017}.

From the full system (electron + phonon) density matrix $\hat \rho$, and following the open quantum system framework~\cite{Petru_book}, the electron and phonon reduced density matrices are $\hat \rho_{\rm e}={\rm Tr}_{\rm ph}\left\{\hat \rho\right\}$ and $\hat \rho_{\rm ph}={\rm Tr}_{\rm e}\left\{\hat \rho\right\}$, where ${\rm Tr}_{\rm e}$ (${\rm Tr}_{\rm ph}$) denotes the trace over the electronic (phononic) degrees of freedom. Their matrix elements, and the electron-phonon correlator that couples them, are $\rho_{12}={\rm Tr}_{\rm e}\{\hat \rho_{\rm e} \hat c^{\dagger}_2\hat c_1\}$, $\rho_{\bf qq'}={\rm Tr}_{\rm ph}\{\hat \rho_{\rm ph}\hat b^{\dagger}_{\bf q'}\hat b_{\bf q}\}$, $\bar \rho_{\bf qq'}={\rm Tr}_{\rm ph}\{\hat \rho_{\rm ph}\hat b_{\bf q'}\hat b_{\bf q}\}$, and $\rho_{12}^{\bf q}={\rm Tr}_{\rm e}{\rm Tr}_{\rm ph}\{\hat \rho \hat c^{\dagger}_2 \hat c_1 \hat b_{\bf q}\}$.

When the phonon system is driven out of equilibrium, the phonon density matrix can be split as $\hat{\rho}_{\rm ph} = \hat{\rho}_{\rm ph}^0 + \delta\hat{\rho}_{\rm ph}$, where $\hat{\rho}_{\rm ph}^0$ is the equilibrium thermal phonon density operator and $\delta\hat{\rho}_{\rm ph}$ its deviation from equilibrium. In general $\delta\hat{\rho}_{\rm ph}$ is an unspecified non-equilibrium superposition of phonon number states, but to study the phonon dynamics we do not need its explicit form: it enters only through the coherent phonon field amplitude $B_{\bf q} = {\rm Tr}_{\rm ph}\left\{\delta\hat{\rho}_{\rm ph}\hat{b}_{\bf q}\right\}$, i.e. the deviation of the phonon amplitude $\hat{b}_{\bf q}$ from its equilibrium value, since ${\rm Tr}_{\rm ph}\{\hat{\rho}_{\rm ph}^0\hat{b}_{\bf q}\}=0$ for a thermalized phonon density matrix. Introducing the fluctuation operator $\delta\hat{b}_{\bf q} = \hat{b}_{\bf q} - B_{\bf q}$, we define the fluctuation fields
\begin{align}\label{Eq:fluct}
    \delta \rho_{\bf q q'}&={\rm Tr}_{\rm ph}\left\{ \hat \rho_{\rm ph} \delta \hat b^{\dagger}_{\bf q'} \delta \hat b_{\bf q} \right\} =\rho_{\bf qq'} - B_{\bf q} B^*_{\bf q'}, \\
    \delta \bar \rho_{\bf q\bf q'}&=\langle \delta \hat b_{\bf q'} \delta \hat b_{\bf q} \rangle=\langle  \hat b_{\bf q'} \hat b_{\bf q} \rangle-B_{\bf q} B_{\bf q'}    , \\
    \delta \rho_{12}^{\bf q}&={\rm Tr}_{\rm e}{\rm Tr}_{\rm ph}\left\{  \hat \rho \hat c^{\dagger}_2 \hat c_1 \hat 
    \delta b_{\bf q}  \right\} = \rho_{12}^{\bf q}- \rho_{12} B_{\bf q}.
\end{align}

Truncating the hierarchy at second order in the electron-phonon matrix element $g_{12}^{\bf q \pm}$ — i.e., decoupling electrons and phonons fields in four-body correlators $\langle \hat c^\dagger \hat c \hat b^\dagger \hat b \rangle$ and neglecting higher order correlations, a detailed derivation is provided in the Supplemental Material~\cite{SM} — closes the system of equations, yielding five coupled EOMs: 
\begin{widetext}
\begingroup
\small
\setlength{\tabcolsep}{4pt}
\setlength{\abovedisplayskip}{1pt}\setlength{\belowdisplayskip}{1pt}
\setlength{\abovedisplayshortskip}{1pt}\setlength{\belowdisplayshortskip}{1pt}
\setlength{\jot}{1.5pt}
\begin{subequations}\label{Eq:EOMset}
\begin{tabular}{|l|l|}
\hline
\text{electron density matrix} &
\begin{minipage}{0.66\linewidth}
\begin{equation}\label{Eq:rho_electron}
\dot \rho_{12} = -i\big[ \Tilde{h}, \rho\big]_{12} + \Big( -i \sum\limits_{\bf q}\sum\limits_{3}(g_{13}^{{\bf q}-}\delta\rho_{32}^{\bf q} + g_{31}^{{\bf q}+}{\delta\rho_{23}^{\bf q}}^*) + \text{H.c.}\Big)
\end{equation}
\end{minipage}
\\ \hline
\text{phonon density matrix} &
\begin{minipage}{0.66\linewidth}
\begin{equation}\label{Eq:delta_rho}
\delta \dot \rho_{\bf qq'} = -i(\omega_{\bf q}-\omega_{\bf q'})\delta\rho_{\bf qq'} -i\sum\limits_{12}\left(-g_{12}^{{\bf q'}-}\delta\rho_{21}^{\bf q} + g_{12}^{{\bf q}+}{\delta\rho_{21}^{\bf q'}}^*\right)
\end{equation}
\begin{equation}\label{Eq:delta_rho_bar}
\delta \dot {\bar \rho}_{\bf q'q} = -i \big(\omega_{\bf q'} + \omega_{\bf q}\big)\delta \bar \rho_{\bf q'q} -i \sum_{12}\big(g_{12}^{{\bf q}+}\delta\rho_{12}^{\bf q'} + g_{12}^{{\bf q'}+}{\delta\rho_{12}^{\bf q}}\big)
\end{equation}
\end{minipage}
\\ \hline
\text{electron-phonon correlation} &
\begin{minipage}{0.66\linewidth}
\begin{equation}\label{Eq:delta_rho12q}
\begin{aligned}
\delta \dot \rho_{12}^{\bf q} =& -i\Big([\Tilde{h},\delta\rho^{\bf q}]_{12} + \omega_{\bf q}\delta\rho_{12}^{\bf q}\Big) -i\sum\limits_{\bf q'}\sum\limits_{34} g_{34}^{{\bf q'}+}(\delta_{\bf qq'}+\delta\rho_{\bf qq'})(\delta_{14}-\rho_{14})\rho_{32} \\
\quad &+ i\sum\limits_{\bf q'}\sum\limits_{34} g_{34}^{{\bf q'}+}\delta\rho_{\bf qq'}(\delta_{32}-\rho_{32})\rho_{14} - i \sum_{\bf q'} \sum_{34} g_{43}^{\bf q' -} \delta \bar \rho_{\bf q' q} \left( \delta_{14}-\rho_{14} \right) \rho_{32} + \\ &+ i \sum_{\bf q'} \sum_{34} g_{43}^{\bf q' -} \delta \bar \rho_{\bf q' q} \left( \delta_{32}-\rho_{32} \right) \rho_{14} 
\end{aligned}
\end{equation}
\end{minipage}
\\ \hline
\text{coherent phonon} &
\begin{minipage}{0.66\linewidth}
\begin{equation}\label{Eq:B}
\dot B_{\bf q}= -i\Big(\omega_{\bf q}B_{\bf q} + \sum\limits_{12} g_{12}^{{\bf q}+}\rho_{12}\Big)
\end{equation}
\end{minipage}
\\ \hline
\end{tabular}
\end{subequations}
\endgroup

\end{widetext}
with
\begin{align}\label{Eq:tildeh}
    \tilde h_{12}(t) &= \varepsilon_1 \delta_{12} + \sum_{\bf q}g_{12}^{{\bf q}-}\big(B_{\bf q} + B_{-{\bf q}}^*\big) + v_{12}^{\rm ext}(t).
\end{align}
%

This set of EOMs is the main result of this letter. It describes the non-Markovian dynamics of coupled electron and phonon fields on an equal footing. It also includes a coherent phonon contribution due to the coupling with $B_{\bf q}$. Eq.~\eqref{Eq:rho_electron} governs the propagation of the electronic density matrix.
Their coupling to the phonon field is mediated by the correlation function in Eq.~\eqref{Eq:delta_rho12q}, which accounts for phonon-induced renormalization and the finite lifetime of electronic excitations.
Eqs.~\eqref{Eq:delta_rho} and~\eqref{Eq:delta_rho_bar} describe, respectively, the number-conserving and anomalous (pair) phonon correlations, both coupled to the electronic subsystem through Eq.~\eqref{Eq:delta_rho12q}. This feedback enables energy and coherence exchange between electrons and phonons beyond a thermal-bath approximation. Finally, Eq.~\eqref{Eq:B} describes the dynamics of coherent phonons, corresponding to lattice distortion associated with a nonzero expectation value of the phonon field. In our formalism, the coherent phonons enter Eqs.~\eqref{Eq:rho_electron} and~\eqref{Eq:delta_rho12q} through the effective Hamiltonian \(\tilde h\). This term captures how coherent lattice motion modifies the evolution of electron-hole pairs, for example by creating polaron structures, and also their influence on the electron-phonon correlation function.

From our formalism it is possible to recover well-established theories. 
By inserting the solution of Eq.~(\ref{Eq:delta_rho12q}) into Eq.~\eqref{Eq:rho_electron} and 
performing the Markovian limit~\cite{Perspective} with a thermal phonon distribution $\rho_{\bf qq'}(t)\simeq \delta_{\bf qq'} n_{\bf q}$, where $n_{\bf q}$ is the phonon occupation, we obtain the Lindblad master equation~\cite{Perspective,Xu21,Rossi2002,Xu2020-qm,Xu2024-lw}.
By then taking the semiclassical diagonal approximation $\rho_{12}=\delta_{12} f_1$, where $f_1$ is the electron occupation, we recover the Boltzmann transport equation~\cite{Xu21,Rossi2002}.

There is also a deep connection between the open quantum dynamics formalism developed here and the NEGF approach. To make this connection explicit, we analytically solve Eqs.~\eqref{Eq:delta_rho12q} and~\eqref{Eq:B} and substitute the results into Eqs.~\eqref{Eq:rho_electron} and~\eqref{Eq:delta_rho}. The coupling between Eq.~\eqref{Eq:B} and Eq.~\eqref{Eq:rho_electron} yields the Ehrenfest, or polaronic, self-energy~\cite{Giu2022,Stef2023}. Eq.~\eqref{Eq:delta_rho12q} is responsible for two distinct contributions: its coupling with Eq.~\eqref{Eq:rho_electron} gives rise to the Fan-Migdal self-energy, while its coupling with Eq.~\eqref{Eq:delta_rho} generates the phonon polarization at the random phase approximation (RPA) level. A detailed derivation is provided in the Supplemental Material. Fig.~\ref{FIG:feynman} schematically shows how the self-energies arise. 
Because our formalism propagates a time-local set of equal-time quantities, its cost is linear in the number of time steps, the same scaling achieved by time-linear Green's function schemes~\mbox{
\cite{Schl2020,Joost2020,Pav2022,PavII2022}}
, reached here directly from the density-matrix hierarchy without invoking the generalized Kadanoff-Baym ansatz\mbox{
\cite{Lip86}}.




%
\begin{figure}
    \includegraphics[width=\linewidth]{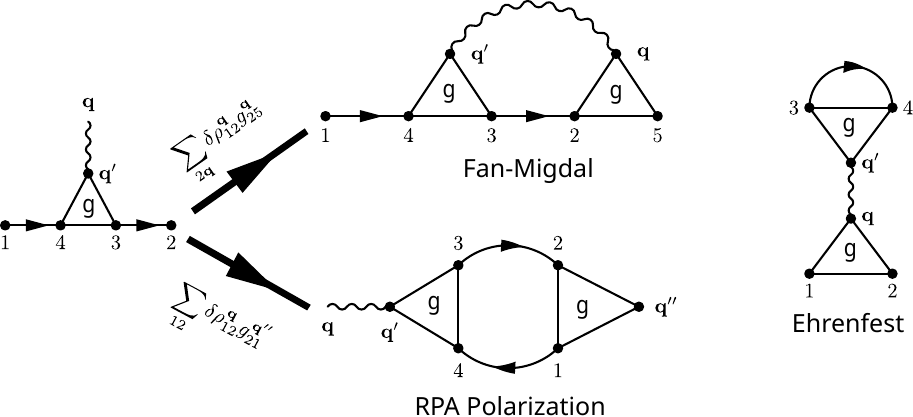}
    \caption{Diagrammatic representation of the electron-phonon correlation function. On the left, the self-energy derived from the coupling of Eq.~\eqref{Eq:delta_rho12q} with Eqs.~\eqref{Eq:rho_electron} and~\eqref{Eq:delta_rho}, showing the link between our theory and the electronic and bosonic collision integrals. On the right, the Ehrenfest self-energy obtained from Eqs.~\eqref{Eq:B} and~\eqref{Eq:tildeh}.} \label{FIG:feynman}
\end{figure}
%


We next test the validity of our set of EOM by comparing against the exact solution of the Holstein dimer. The Hamiltonian of the model is 
\begin{align}\label{Eq:HolsteinH}
    \hat H^{H}(t)=\hat H_0+\hat H^{\rm int}+\hat H^{\rm ext}(t)
\end{align}
where
\begin{align}
    &\hat H_0=-t_{\rm hop}\sum_{\sigma,<i,j>}\hat c_{i\sigma}^\dagger \hat c_{j\sigma} + \omega_{\rm ph} \hat b^{\dagger} \hat b \label{Eq:H0} \\
    &\hat H^{\rm int}= \frac{g}{\sqrt{2}}\left(\hat b^{\dagger}+ \hat b\right)\left(\hat n_1-\hat n_2\right) \label{Eq:H_int} \\
    &\hat H^{\rm ext}(t)=\sum_{i}v_i \hat n_i \text{sin}\left(t \frac{\pi}{4}\right) e^{-\frac{t^2}{2T_{\rm p}^2}}. \label{Eq:H_ext}
\end{align}
The parameter $t_{\rm hop}$ and $\omega_{\rm ph}$ describe the kinetic electron hopping energy and phonon frequency, respectively. $i$ and $j$ refer to the sites while $\sigma$ to the spin. $g$ is the e-ph coupling parameter, obtained from the e-ph tensor with the relation $g_{\alpha \beta}^{\bf q}=\frac{g}{\sqrt{2}} \delta_{1,\bf{q}} \delta_{\sigma_{\alpha} \sigma_{\beta}} \delta_{\alpha\beta}(\delta_{\alpha,1}-\delta_{\beta,2})$~\cite{Robert2015,Robert2021}. $\hat n_1$ and $\hat n_2$ are the electronic density operator of the two sites.
The last contribution is a time-dependent external field that we use to drive the system out of equilibrium. We use an oscillatory Gaussian-damped perturbation with $T_{\rm p}=t_{\rm hop}^{-1}$ that interacts with the electronic density $\hat n_i$. The amplitude of the external field is given by $v_i$. We assume a symmetric dimer with $v_1=-v_2 \equiv v$. The model is described by two dimensionless parameters,
\begin{align}
    \gamma= \frac{\omega_{\rm ph}}{t_{\rm hop}} && \lambda= \frac{2g^2}{t_{\rm hop} \omega_{\rm ph}},
\end{align}
describing the ratio between phonon and electron energies and the coupling strength between them, respectively. 


The dimer is symmetric with respect to spin. Therefore, it is possible to study the 1-RDM for only spin up. In this way, the 1-RDM for the dimer is simply a $2\times 2$ matrix. Under this symmetry, the system of equations~\eqref{Eq:rho_electron}-\eqref{Eq:B} has to be modified. We provide the new set of equations and their derivation in the Supplemental Material.

To obtain the exact solution as a reference, we perform the time evolution of the full density matrix; we then evaluate the reduced density matrices defined above to obtain the exact solution of 1-RDM. To ensure convergence, the phonon space is truncated at a maximum occupation number of $30$. 
All simulations employ a RK4 algorithm to solve the dynamics. 


%
\begin{figure}
    \includegraphics[width=\linewidth]{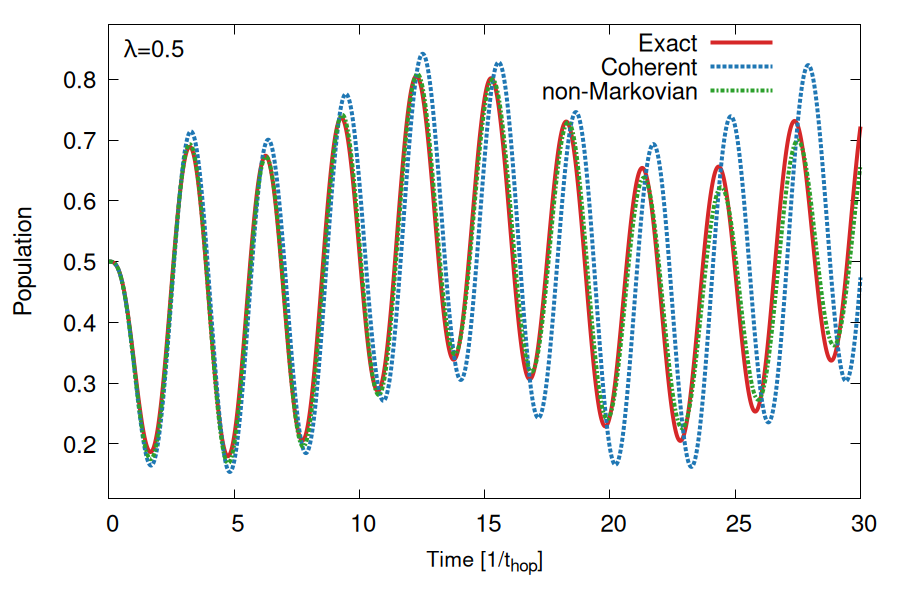}
    \caption{Dynamics of the occupation of the first site for different level of theories. Exact dynamics (solid red curve), coherent phonons dynamics (dotted blue curve), and non-Markovian dynamics (dot-dashed green curve). The dynamics is obtained with $\lambda=0.5$ and $\gamma=0.5$} \label{FIG:dynamics}
\end{figure}
\begin{figure*}[t]
    \includegraphics[width=\linewidth]{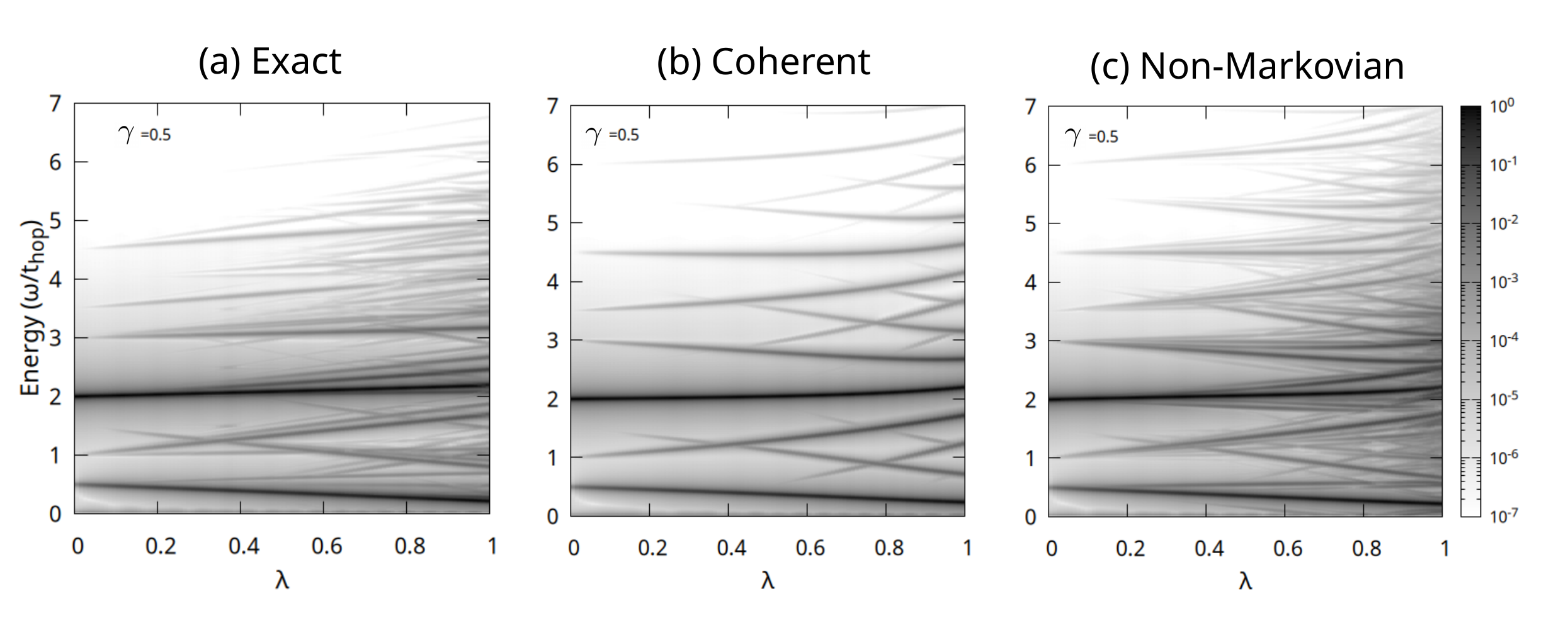}
    \caption{Spectra as a function of the dimensionless coupling constant $\lambda$. (a) Exact result. (b) Coupled Eqs.~\eqref{Eq:rho_electron} and~\eqref{Eq:B}. (c) Solution of the full set of non-Markovian equations~\eqref{Eq:rho_electron}-\eqref{Eq:B}. The Non-Markovian result shows a broadening of the peaks similarly to the exact result, the coherent phonon-electron dynamics do not capture this feature.}\label{FIG:spectral}
\end{figure*}
We test two different approximations. We refer to the first as ``coherent dynamics," the solution of Eqs.~\eqref{Eq:rho_electron} and~\eqref{Eq:B}, and the second as ``non-Markovian dynamics," the solution of the four-EOM system Eqs.~\eqref{Eq:rho_electron}--\eqref{Eq:B}, without including Eq.~\eqref{Eq:delta_rho_bar}. The test with the full five-EOM system is reported in the Supplemental Material.
We initialize our simulations at the fixed point of the EOMs that minimizes the total energy, obtained by first finding the roots of the right-hand side of the system and then selecting the one with the lowest energy. We drive the system out of equilibrium through an external perturbation, described by Eq.~\eqref{Eq:H_ext}. To describe the interaction with a strong external field we set its amplitude to be equal to the hopping energy $v=t_{\rm hop}$. Fig.~\ref{FIG:dynamics} shows the population of the first site as a function of time for exact, non-Markovian and coherent dynamics at $\lambda=0.5$ and $\gamma=0.5$. We observe that both approximations reproduce the overall dynamics, but the non-Markovian dynamics yields closer agreement with the exact results.
%

%

%
We run the dynamics until a final time $T_{\rm f}=400t_{\rm hop}^{-1}$. Fig.~\ref{FIG:spectral} shows the spectra obtained from the Fourier transform of the polarization~\cite{Tak2007,Att2011,Luppi2010,Per2015},
\begin{equation}
    P(t)=\frac{1}{2}\left(\rho_{22}(t)-\rho_{11}(t)\right)
\end{equation}
as a function of the dimensionless coupling constant $\gamma$ and $\lambda$, respectively.
Spectra are obtained from the discrete Fourier transform of P(t), apodized with a Hann window to suppress spectral leakage from the finite time window $[0, T_f]$~\cite{Numerical}.

Fig.~\ref{FIG:spectral} shows the rich spectra obtained due to the strong external field we apply. We can recognize two main contributions. The peak at $E=2$ represents the quasiparticle excitation (electron-hole pair), while the peak at $E=0.5$ is the polaronic contribution, i.e. the renormalized excitation energy produced by the Ehrenfest (polaron) self-energy of Eq.~\eqref{Eq:B}. These two main excitations are well described by both approximations. The non-Markovian spectra show a broadening of the excitations absent in the coherent dynamics; this effect is most visible for large values of $\lambda$. As $\lambda$ increases, both the exact and non-Markovian spectra develop a rich multi-peak structure. The absence of these features in the coherent-dynamics solution shows that they cannot be attributed solely to a mean-field coherent lattice displacement. 
Instead, they arise from the electron–phonon correlations encoded in Eq.~\eqref{Eq:delta_rho12q}, i.e. the Fan–Migdal self-energy, including multi-phonon processes, and are captured only once this electron–phonon correlation EOM is propagated.
%

%

%

Energy is conserved throughout the dynamics, and each energy contribution (electron, phonon, and electron-phonon) agrees closely with the exact solution; this check is presented in the End Matter.

In summary, we have presented a non-Markovian open quantum systems formalism for
nonequilibrium electron-phonon coupled dynamics, based on a closed set of five equations
of motion for the electronic 1-RDM, the phonon density matrix, the anomalous phonon
density matrix, the coherent phonon amplitude, and the electron-phonon correlations,
obtained directly from the open-system density matrix without recourse to the generalized
Kadanoff-Baym ansatz. The formalism recovers the Fan-Migdal, RPA polarization, and
Ehrenfest self-energies, and reduces to the Lindblad and Boltzmann equations in the
appropriate limits, establishing its consistency with established theories. Because only
equal-time quantities are propagated, its cost is linear in the number of time steps.
Benchmarked against the exact solution of the Holstein dimer under strong external
perturbation, the formalism separates two physically distinct contributions to the
spectrum: the coherent polaronic peak, produced by the Ehrenfest channel of
Eq.~\eqref{Eq:B}, and the finite linewidths and multiphonon structure, produced by the
Fan--Migdal correlations of Eq.~\eqref{Eq:delta_rho12q} and absent from the coherent
dynamics; energy is conserved channel by channel throughout.
This work provides a unified and computationally tractable framework for the study of
ultrafast carrier relaxation, polaron formation, and nonequilibrium phonon dynamics in
strongly driven systems. Here the Holstein dimer serves as an exact benchmark for the
equations of motion; because the \textit{ab-initio} electron--phonon coupling
infrastructure that enters them is already available~\cite{Xu2024-cb,Xu21,lee_electronphonon_2023},
the formalism can be applied directly to driven real materials, which we are currently
pursuing.

\textbf{Acknowledgments} 
We thank Ilya A Esterlis and Gianluca Stefanucci for very helpful discussions. This work is primarily supported by the Computational Chemical Sciences program within the Office of Science of the DOE under Grant No. DE-SC0023301. Part of the work is supported by  the National Science Foundation
under Grant No. DMR-2143233. Calculations were carried out at the National Energy Research Scientific Computing Center (NERSC), a U.S. Department of Energy Office of Science User Facility operated under Contract No. DEAC02-05CH11231.
\bibliography{main}

\clearpage
\onecolumngrid
\begin{center}
  \textbf{\large End Matter}
\end{center}
\twocolumngrid
\vspace{6pt}

\section*{Appendix: Energy conservation}

\begin{figure}[t]
    \includegraphics[width=\linewidth]{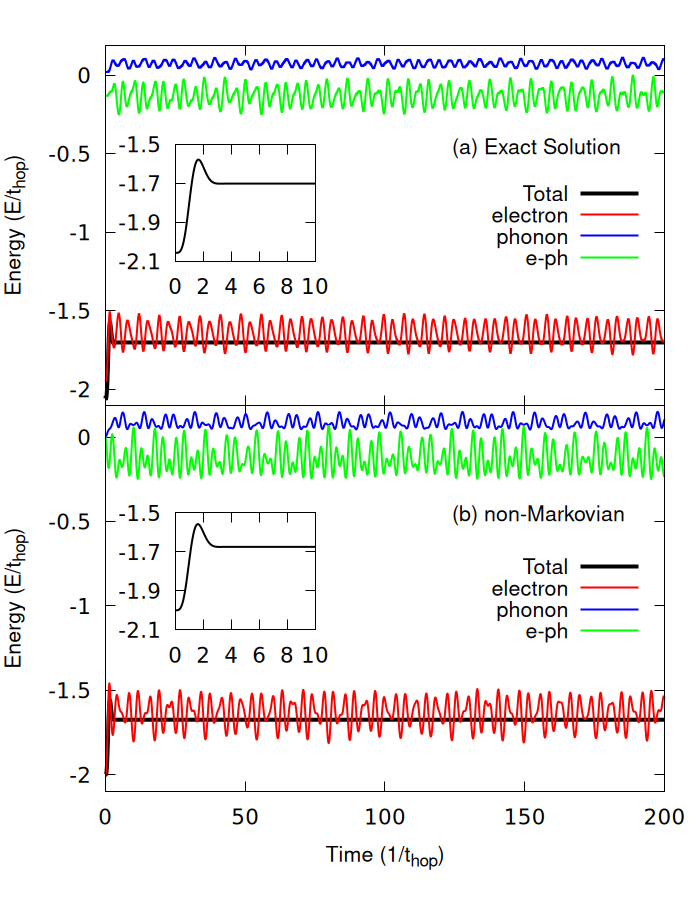}
    \caption{Comparison among different energy contributions in Eq.~\eqref{Eq:energy}. Black curve, total energy. Red, blue, and green line, energy contributions of the electron, phonon, and electron-phonon correlation, respectively. Top panel, exact result. Bottom panel, non-Markovian dynamics. Insets, zoom on the variation of the total energy due to the external pump. This plot is obtained for $\lambda=0.5$ and $\gamma=0.5$.} \label{FIG:energy}
\end{figure}
An important physical condition that has to be verified is energy conservation. By averaging the time-independent part of the Hamiltonian in Eq.~\eqref{Eq:HolsteinH} we obtain

\begin{align}\label{Eq:energy}
    E =& \underbrace{\sum_1 \epsilon_1 \rho_{11}}_{\text{electron}}+ \underbrace{\sum_{\bf q} \omega_{\bf q} (B_{\bf q}B_{\bf q}^*+\delta\rho_{\bf qq})}_{\text {phonon}} + \nonumber \\ +& \underbrace{\sum_{12\bf q} g_{12}^{\bf q -}(B_{\bf q} \rho_{21}+\delta\rho_{21}^{\bf q})+\sum_{12 \bf q} g_{12}^{\bf q +}(B^*_{\bf q} \rho_{12} +\delta \rho_{21}^{\bf q*})}_{\text {electron-phonon}} .
\end{align}
The energy depends only on the computed dynamical quantities, so it is easy to calculate for both exact and non-Markovian dynamics. Fig.~(\ref{FIG:energy}) shows the comparison between the exact and non-Markovian solution for $\lambda=0.5$ and $\gamma=0.5$. We observe that, after an initial total-energy variation due to the external pump, the total energy remains constant. Moreover, each energy contribution (electron, phonon, and electron-phonon) is in close agreement with the exact solution. We further verified that the total electron number, ${\rm Tr}\left\{\rho\right\}$, is conserved throughout the dynamics.
%

\end{document}